\documentclass[showpacs,aps,pre,superscriptaddress]{revtex4}
\usepackage{graphicx}
\usepackage{amsfonts}
\usepackage{color, verbatim}

\begin{document}
\title{ Existence, Stability \& Dynamics of Nonlinear Modes in a 2d Partially $\mathcal{PT}$ Symmetric Potential}
\author{J. D'Ambroise}
\affiliation{ Department of Mathematics, Computer \& Information Science, State University of New York (SUNY) College at Old Westbury, Westbury, NY, 11568, USA; dambroisej@oldwestbury.edu}
\author{P.G. Kevrekidis}
\affiliation{Department of Mathematics and Statistics, University of Massachusetts,
Amherst, MA, 01003, USA; kevrekid@math.umass.edu}

\begin{abstract}
It is known that multidimensional complex potentials obeying $\mathcal{PT}$-symmetry may possess all real spectra and continuous families of solitons. Recently it was shown that for multi-dimensional systems these features can persist when the parity symmetry condition is relaxed so that the potential is invariant under reflection in only a single spatial direction.  We examine the existence, stability
  and dynamical properties of localized modes within the cubic nonlinear Schr\"odinger equation in such a scenario of partially $\mathcal{PT}$-symmetric potential.
\end{abstract}

\maketitle

\section{Introduction}

The study of $\mathcal{PT}$ (parity--time) symmetric systems was initiated
through the works of Bender and collaborators~\cite{Bender1,Bender2}.
Originally, it was proposed as an alternative to
the standard quantum theory, where the Hamiltonian is postulated to be
Hermitian. In these works, it was instead found that
Hamiltonians invariant under $\mathcal{PT}$-symmetry, which are not necessarily Hermitian, may still give
rise to completely real spectra. Thus, the proposal of Bender and
co-authors was that these Hamiltonians are appropriate for the description
of physical settings. In the important case
of  Schr{\"{o}}dinger-type Hamiltonians,
which include the usual kinetic-energy operator and the potential term, $%
V(x) $, the $\mathcal{PT}$-invariance is consonant with
complex potentials, subject to the constraint that $V^{\ast }(x)=V(-x)$.

A decade later, it was realized (and since then it has led to a decade
of particularly fruitful research efforts) that this idea can find
fertile ground for its experimental realization although not in quantum
mechanics where it was originally conceived. In this vein, numerous
experimental realizations sprang up in the areas of linear and
nonlinear optics~\cite{Ruter,Peng2014,peng2014b,RevPT,Konotop},
electronic circuits
\cite{Schindler1,Schindler2,Factor}, and mechanical systems \cite{Bender3},
among others. Very recently, this now mature field of research
has been summarized in two comprehensive reviews~\cite{RevPT,Konotop}.

One of the particularly relevant playgrounds for the exploration
of the implications of $\mathcal{PT}$-symmetry is that of nonlinear
optics, especially because it can controllably involve the interplay
of  $\mathcal{PT}$-symmetry and nonlinearity.
In this context, the propagation of light (in systems such
as optical fibers or waveguides~\cite{RevPT,Konotop})
is modeled by the nonlinear Schr\"odinger equation of the form:
\begin{eqnarray}
\label{nls}
i\Psi_z + \Psi_{xx} + \Psi_{yy} +  U(x,y)\Psi + \sigma |\Psi|^2\Psi = 0.
\end{eqnarray}
In the optics notation that we use here, the evolution direction is denoted by $z$, the propagation distance.
Here, we restrict our considerations to two spatial dimensions and
assume that the potential $U(x,y)$ is complex valued, representing
gain and loss in the optical medium, depending on the sign of the imaginary part (negative for gain, positive for loss) of the potential.
In this two-dimensional setting, the condition of full $\mathcal{PT}$-symmetry in two dimensions is that $U^*(x,y) = U(-x,-y)$.  Potentials with full $\mathcal{PT}$ symmetry have been shown to support continuous families of soliton solutions \cite{OptSolPT,WangWang,LuZhang,StabAnPT,ricardo}.
However, an important recent development was the fact that the condition of
(full) $\mathcal{PT}$ symmetry can be relaxed.  That is, either the condition $U^*(x,y)=U(-x,y)$ or $U^*(x,y)=U(x,-y)$ of, so-called,
partial $\mathcal{PT}$ symmetry can be imposed, yet the
system will still maintain all real spectra and continuous families of soliton solutions \cite{JYppt}.  

In the original contribution of~\cite{JYppt}, only the focusing nonlinearity
case was considered for two select branches of solutions and the stability
of these branches was presented for isolated parametric cases (of the
frequency parameter of the solution). Our aim in the present work
is to provide a considerably more ``spherical'' perspective of the
problem. In particular, we examine the bifurcation of nonlinear
modes from {\it all three}
point spectrum eigenvalues of the underlying linear Schr{\"o}dinger
operator of the partially $\mathcal{PT}$-symmetric potential.
Upon presenting the relevant model (section 2),
we perform the relevant continuations (section 3) unveiling the existence
of nonlinear branches {\it both} for the focusing and for the
defocusing nonlinearity case. We also provide a systematic view
towards the stability of the relevant modes (section 4),
by characterizing their
principal unstable eigenvalues as a function of the intrinsic
frequency parameter of the solution. In section 5, we complement our
existence and stability analysis by virtue of direct numerical
simulations that manifest the result of the solutions' dynamical instability
when they are found to be unstable. Finally, in section 6,
we summarize our findings and present our conclusions, as well
as discuss some possibilities for future studies.

\section{Model, Theoretical Setup and Linear Limit}

Motivated by the partially $\mathcal{PT}$-symmetric
setting of~\cite{JYppt}, we consider the complex potential $U(x,y) = V(x,y) + iW(x,y)$ where
\begin{eqnarray} 
 V &=& \left( ae^{ - (y - y_0)^2} + be^{ - (y + y_0)^2}\right)\left(e^{-(x - x_0)^2 } + e^{-(x + x_0)^2 }\right)\nonumber
\\
W &=& \beta\left(ce^{ - (y - y_0)^2} + de^{ - (y + y_0)^2}\right)\left(e^{-(x - x_0)^2 } - e^{-(x + x_0)^2 }\right).\label{VW}
\end{eqnarray}
with real constants $\beta$, $a\neq b $ and $c\neq -d$.  The potential is chosen with partial $\mathcal{PT}$-symmetry so that $U^*(x,y)=U(-x,y)$.  That is, the real part is even in the $x$-direction with $V(x,y) = V(-x,y)$ and the imaginary part is odd in the $x$-direction with $-W(x,y) = W(-x,y)$.    The constants $a,b,c,d$ are chosen such that there is no symmetry in the $y$ direction.  

\begin{figure}
\centering
\includegraphics[width=3.5in]{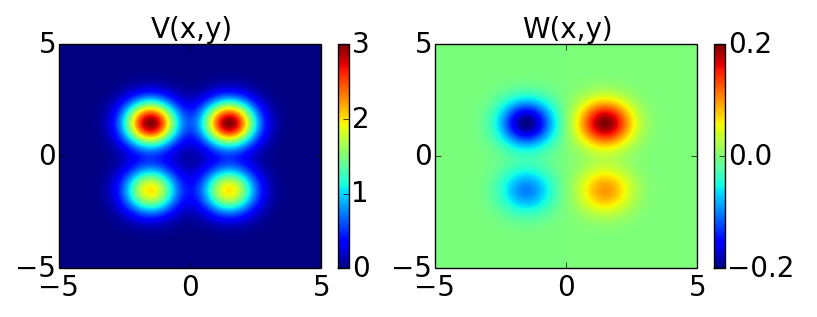}
\caption{The plots show the spatial distribution of real ($V$,
  left panel) and imaginary ($W$, right panel) parts of the potential $U$ with $x_0=y_0=1.5$. }
\label{figVW}
\end{figure}

\begin{figure}
\centering
\includegraphics[width=3.5in]{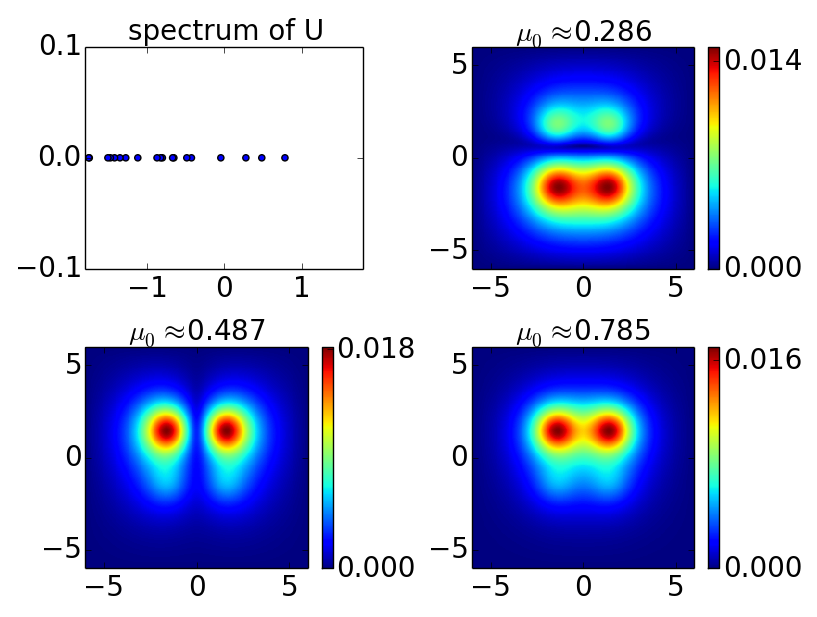}
\caption{ The top left plot shows the spectrum of Schr{\"o}dinger
  operator associated with the potential $U$ in the complex plane (see
  also the text).   Plots of the magnitude of the normalized eigenvectors for the three discrete eigenvalues $\mu_0$ are shown in the other three plots.}
\label{figVWeigs}
\end{figure}

In \cite{JYppt} it is shown that the spectrum of the potential $U$ can be all real as long as $|\beta|$ is below a threshold value, after which a
($\mathcal{PT}$-) phase transition occurs; this is a standard property of $\mathcal{PT}$-symmetric potentials.  We focus on the case $\beta=0.1$ and $a=3, b=c=2, d=1$ for which the spectrum is real, i.e., below the relevant transition threshold.   Figure~\ref{figVW} shows plots of the potential $U$.
The real part of the potential is shown on the left, while the
imaginary part associated with gain-loss is on the right;
the gain part of the potential corresponds to $W<0$ and occurs
for $x<0$, while the loss part with $W>0$ occurs for $x>0$.  Figure~\ref{figVWeigs}
shows  the spectrum of $U$, i.e., eigenvalues for the underlying
linear Schr{\"o}dinger 
problem $(\nabla^2 + U)\psi_0 = \mu_0\psi_0$.  The figure also shows the corresponding eigenvectors for the three discrete real eigenvalues $\mu_0$. It is
from these modes that we will seek bifurcations of nonlinear
solutions in what follows.

\section{Existence: Nonlinear Modes Bifurcating from the Linear Limit}

As is customary, we focus on stationary soliton solutions of (\ref{nls}) of the form $\Psi(x,y,t) = \psi(x,y)e^{i\mu z}$.  Thus one obtains the following stationary equation for $\psi(x,y)$.
\begin{equation}
\psi_{xx} + \psi_{yy} + U(x,y)\psi + \sigma|\psi|^2\psi = \mu \psi
\label{stateq}
\end{equation}
In \cite{JYppt} it is discussed that a continuous family of solitons bifurcates from each of the linear solutions in the presence of nonlinearity.  
In order to see this let $\mu_0$ be a discrete simple real eigenvalue of the potential $U$ (such as one of the positive real eigenvalues in the top left plot of Fig. \ref{figVWeigs}).  Now, following~\cite{JYppt}, expand $\psi(x,y)$ in terms of $\epsilon = |\mu-\mu_0| << 1$ and substitute the expression
\begin{equation}
\psi(x,y) = \epsilon^{1/2}\left[ c_0 \psi_0 + \epsilon \psi_1 + \epsilon^2 \psi_2  + \dots \right]
\label{epsexpand}
\end{equation}
into equation (\ref{stateq}).  This gives the equation for $\psi_1$ as
\begin{equation}
L\psi_1 = c_0\left( \rho\psi_0 - \sigma |c_0|^2|\psi_0|^2\psi_0 \right)
\label{psi1eq}
\end{equation}
where $\rho = {\rm sgn}(\mu-\mu_0)$ and 
\begin{equation}
|c_0|^2 = \frac{\rho \langle \psi_0^*, \psi_0 \rangle}{\sigma \langle \psi_0^*, |\psi_0|^2 \psi_0\rangle}.
\label{c0}
\end{equation}
Here,  $\psi_0^*$ plays the role of the adjoint solution to $\psi_0$.

Thus in order to find solutions of (\ref{stateq}) for $\sigma = \pm 1$ we perform a Newton continuation in the parameter $\mu$ where the initial guess for $\psi$ is given by the first two terms of (\ref{epsexpand}).  The bottom left panel of Figure \ref{muvpow} shows how the (optical) power $P(\mu) = \int\int |\psi|^2 dxdy$  of the solution grows as a function of increasing $\mu$ for $\sigma=1$, or as a function of decreasing $\mu$ for $\sigma=-1$ (from the linear
limit).  The first branch begins at the first real eigenvalue of $U$ at $\mu_0 \approx 0.286$, the second branch at $\mu_0 \approx 0.487$, and the third branch begins at $\mu_0 \approx 0.785$.   Plots of the solutions and their corresponding time evolution and stability properties are shown in the next section.

As a general starting point comment for the properties of the branches,
we point out that all the branches populate both the gain and the
loss side. In the branch starting from $\mu_0 \approx 0.286$, all 4 ``wells'' of
the potential of Fig.~\ref{figVW} appear to be populated, with the
lower intensity ``nodes'' being more populated and the higher intensity
ones less populated. The second branch starting at $\mu_0 \approx 0.487$,
as highlighted also in~\cite{JYppt},
possesses an anti-symmetric structure in $x$ (hence the
apparent vanishing of the density at the $x=0$ line). Both in the second
and in the third branch, the higher intensity nodes of the potential
appear to bear a higher intensity.

\section{Stability of the Nonlinear Modes: Spectral Analysis}

\begin{figure}
\centering
\hspace{-.2in}\includegraphics[width=3.5in]{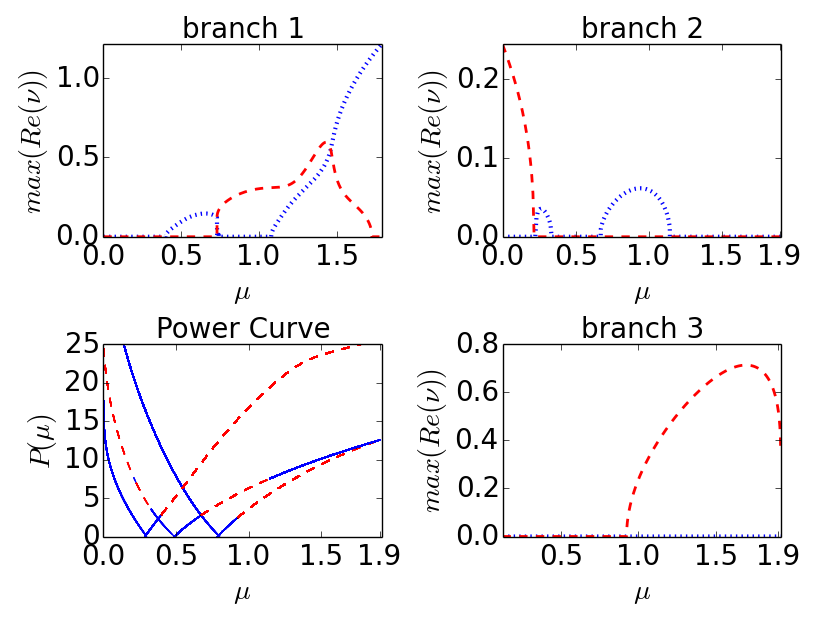}
\caption{The bottom left plot shows the power of the solution $\psi$ plotted in terms of the continuation parameter $\mu$.  The curves begin at the lowest power (i.e., at the linear limit) at the discrete real eigenvalues of approximately $0.286$ (branch 1), $0.487$ (branch 2), $0.785$ (branch 3).  Each power curve is drawn with its corresponding stability noted:  a blue solid curve denotes a stable solution and a red dashed curve denotes an unstable solution.  The other three plots track the maximum real part of eigenvalues $\nu$ as a function of the continuation parameter $\mu$:  the red dashed line represents the max real part of eigenvalues that are real (exponential instability)
  while the blue dotted line tracks the max real part for eigenvalues that have nonzero imaginary part (quartets); this case corresponds to oscillatory
  instabilities. }
\label{muvpow}
\end{figure}

The natural next step is to identify the stability of the solutions.
This is predicted by using the linearization ansatz:
\begin{equation}
\label{pert}
\Psi = e^{i\mu z}\left( \psi + \delta \left[ a(x,y)e^{\nu z} + b^*(x,y)e^{\nu^*z}\right] \right)
\end{equation}
which yields the order $\delta$ linear system
\begin{equation}
\left[ \begin{array}{cc} M_1 & M_2 \\ -M_2^* & -M_1^* \end{array} \right] \left[ \begin{array}{c} a\\ b \end{array} \right] = -i \nu \left[ \begin{array}{c} a\\ b \end{array} \right]
\end{equation}
where $M_1 = \nabla^2 + U - \mu + 2\sigma|\psi|^2$, $M_2 = \sigma \psi^2$.  Thus ${\rm max}({\rm Re}(\nu)) > 0$ corresponds to instability and ${\rm max}({\rm Re}(\nu)) = 0$ corresponds to (neutral) stability.

\begin{figure}
\centering
\includegraphics[width=3.5in]{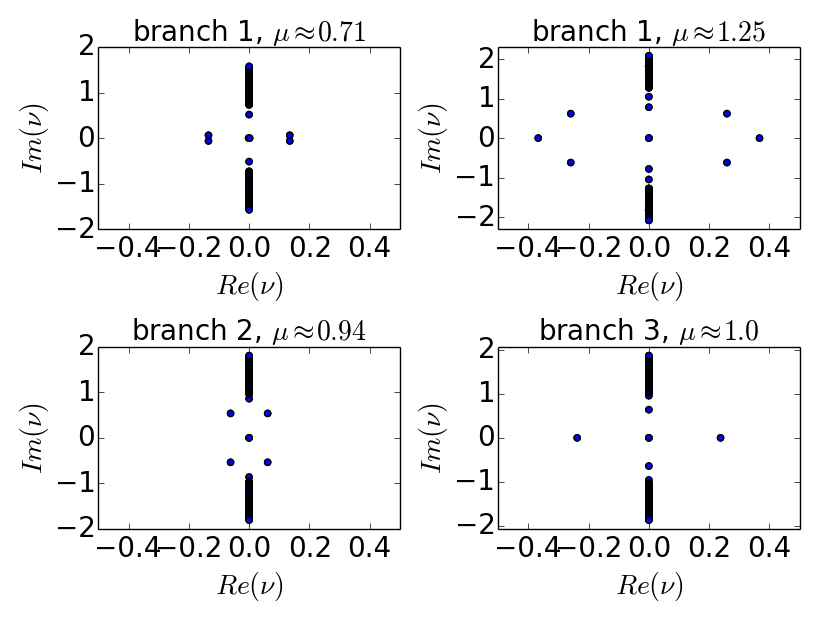}
\caption{Eigenvalues are plotted in the complex plane
  $(Re(\nu),Im(\nu))$ for a few representative solutions.  One can compare
  the maximal real part with Figure \ref{muvpow}.  For example, the top left complex plane plot here shows that for branch 1 at $\mu\approx 0.71$ the eigenvalues with maximum real part are complex; this agrees with the top left plot of Figure \ref{muvpow} where at $\mu\approx 0.71$ the blue dotted curve representing complex eigenvalues is bigger.  Similarly one can check the other three eigenvalue plots here also agree with what is shown in Figure \ref{muvpow}, the top right for branch 1,
  the bottom left for branch 2 and the bottom right for
  branch 3.}
\label{eigs}
\end{figure}

In the bottom left panel of Figure \ref{muvpow} the power curve is drawn with stability and instability as determined by $\nu$ noted by the solid or dashed curve, respectively.  The other three plots in Figure \ref{muvpow} show the maximum real part of eigenvalues $\nu$ for each of the three branches:  the red dashed curve is the max real part of real eigenvalue pairs, and the blue dotted curve is the max real part of eigenvalue quartets with nonzero imaginary part.
The former corresponds to exponential instabilities associated with
pure growth, while the latter indicate so-called oscillatory instabilities,
where growth is present concurrently with oscillations.
In Figure \ref{eigs} we plot some example eigenvalues in the complex plane for some sample unstable solutions. The dominant unstable eigenvalues within
these can be seen to be consonant with the growth rates reported
for the respective branches (and for these parameter values) in
Fig.~\ref{muvpow}.

The overarching conclusions from this stability analysis are as follows.
The lowest $\mu$ branch, being the ground state in the defocusing
case, is always stable in the presence of the self-defocusing nonlinearity.
For the parameters considered, generic stability is also
prescribed for the third branch under self-defocusing nonlinearity.
The middle branch has a narrow interval of stability and then becomes
unstable, initially (as shown in the top right of Fig.~\ref{muvpow})
via an oscillatory instability and then through an exponential one. In
the focusing case (that was also focused on in~\cite{JYppt}
for the second and third branch), all three branches appear
to be stable immediately upon their bifurcation from the linear
limit, yet, all three of them subsequently become unstable.
Branch 1 (that was not analyzed previously)
features a combination of oscillatory and exponential
instabilities. Branch 2 features an oscillatory instability which,
however, only arises for a finite interval of frequencies $\mu$,
and the branch restabilizes. On the other hand, branch 3, when it
becomes unstable it is through a real pair. Branches 2 and 3
terminate in a saddle-center bifurcation near
$\mu=1.9$. The eigenvalue panels of Fig.~\ref{eigs} confirm
that the top panels of branch 1 may possess one or two concurrent
types of instability (in the focusing case), branch 2 (bottom left)
can only be oscillatorily unstable in the focusing case
(yet as is shown in Fig.~\ref{muvpow} it can feature both types
of
instabilities in the defocusing case), while
branch 3, when unstable in the focusing case is so via a real eigenvalue pair.

\section{Dynamics of Unstable Solutions}
 
\begin{figure}
\centering
\includegraphics[width=3.5in]{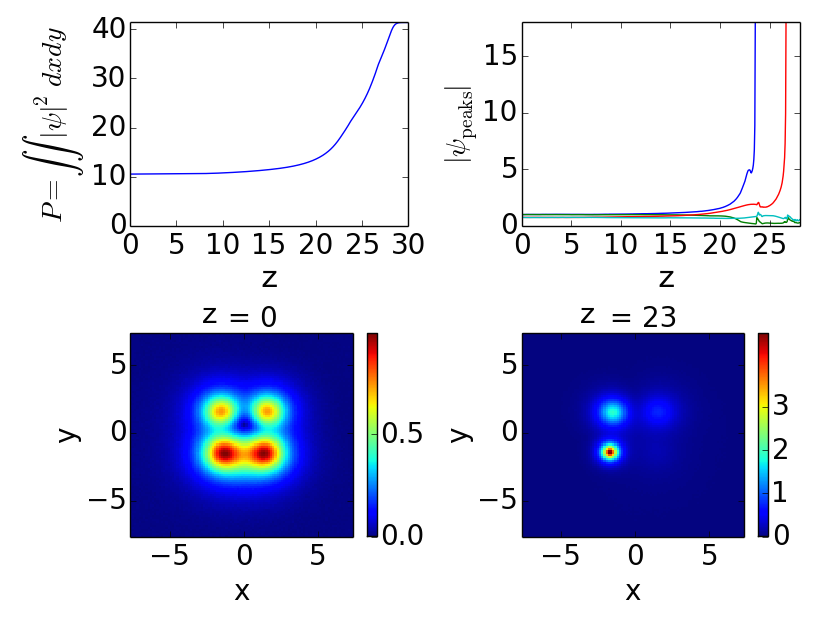}
\caption{This figure shows the time evolution of the branch 1 solution for the value $\mu\approx 0.71$.   The bottom left plot shows the magnitude of the solution $|\Psi|$ at $t=0$.   Observe that this solution has four peaks in its magnitude over the two-dimensional spatial grid.  The bottom right plot shows the solution at $t=23$.  Observe that the magnitudes of the peaks on the left side have increased.  The top left plot shows the time evolution of the power of the solution as a function of the evolution variable $z$.
  The top right plot here shows the evolution of the four peaks in the magnitude of the solution as a function of $z$ (blue = bottom left peak, red = top left peak, green = bottom right peak, cyan = top right peak).
}
\label{rkp1}
\end{figure}

\begin{figure}
\centering
\includegraphics[width=3.5in]{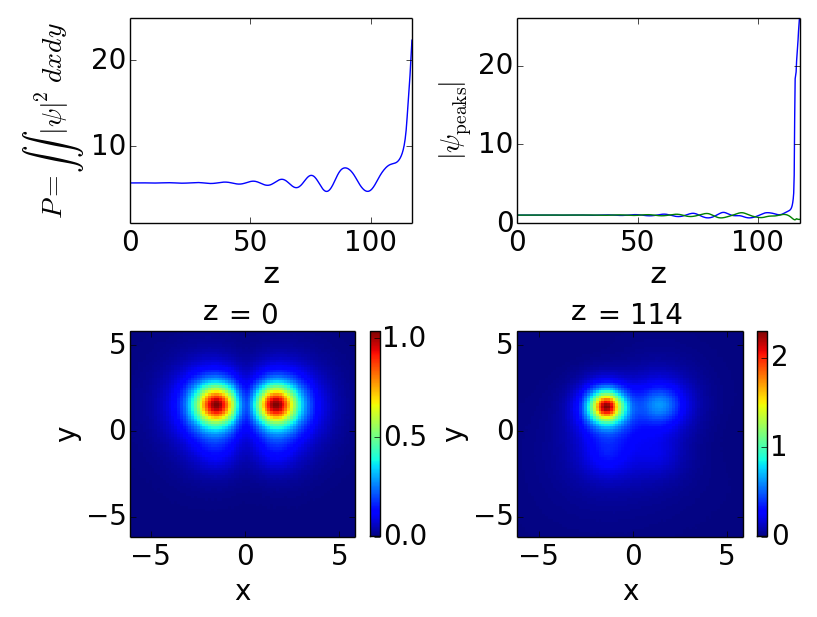}
\caption{This figure is similar to Figure \ref{rkp1} (the final evolution
  distance however is about $z=114$).  Here the plots correspond to the time evolution of the branch 2 solution for the value $\mu\approx 0.94$.  In the top right plot, the blue  curve corresponds to the left peak of the magnitude of the solution over $z$ and the green corresponds to the right peak of the magnitude.
}
\label{rkp2}
\end{figure}

\begin{figure}
\centering
\includegraphics[width=3.5in]{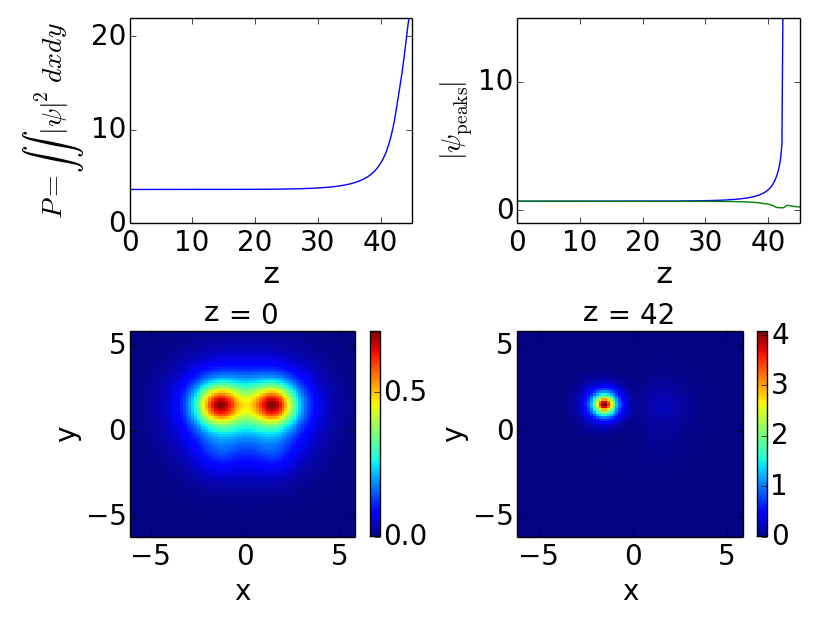}
\caption{This figure is similar to Figure \ref{rkp1} (with an evolution
  up to distance $z=42$).  Here, the plots correspond to the
  evolution of the branch 3 solution for the value $\mu\approx 1.0$.    In the top right plot, the blue curve corresponds to the left peak of the magnitude of the solution over $z$ and green corresponds to the right peak of the magnitude.
  Clearly, once again, the gain side of the solution eventually dominates.
}
\label{rkp3}
\end{figure}

Figures \ref{rkp1}, \ref{rkp2}, \ref{rkp3} show the time evolution of three unstable solutions, one on each branch.  All three time evolution examples we show here have a value of $\sigma = 1$.  That is, they each correspond to a mu-value that is bigger than the initial discrete value $\mu_0$
and pertain to the focusing case.  The time evolution figures show a similar feature for the unstable solutions, namely that over time the magnitude of the solutions will increase on the left side of the spatial grid.  This agrees with what is expected from $\mathcal{PT}$-symmetry since the left side of the spatial grid corresponds to the gain side of the potential $U$. Importantly, also,
the nature of the instabilities varies from case to case,
and is consonant with our stability expectations based on the results of
the previous section.

In  Fig.~\ref{rkp1} branch 1
(for the relevant value of the parameter $\mu$) features
an oscillatory instability (but with a small imaginary part).
In line with this, we observe
a growth that is principally exponential (cf. also the
top panel for the power of the solution), yet features also some
oscillation in the amplitude of the individual peaks. It should be noted
here that although two peaks result in growth and
two in decay (as expected by the nature of $W$ in this case),
one of them clearly dominates between the relevant amplitudes.

In Fig.~\ref{rkp2}, it can be seen that branch 2, when unstable in the
focusing case, is subject to an oscillatory instability (with a
fairly significant imaginary part). Hence the growth
is not pure, but is accompanied by oscillations as is clearly visible
in the top left panel. In this case, among the two principal peaks
of the solution of branch 2, only the left one (associated with the
gain side) is populated after the evolution shown.

Lastly, in branch 3, the evolution (up to $z=42$) manifests the existence
of an exponential instability. The latter leads once again the gain
part of the solution to indefinite growth resulting in one of
the associated peaks growing while the other (for $x>0$ on the lossy side)
featuring decay.

It is worthwhile to mention that in the case of branch 2, the only
branch that was found (via our eigenvalue calculations) to be unstable
in the self-defocusing case, we also attempted to perform
dynamical simulations for $\sigma=-1$. Nevertheless, in all the cases
considered it was found that, fueled by the defocusing nature of
the nonlinearity, a rapid spreading of the solution would
take place (as $z$ increased), leading to a rapid interference of
the results with the domain boundaries. For that reason, this
evolution is not shown here.


\section{Conclusions \& Future Challenges}

In the present work, we have revisited the partially $\mathcal{PT}$-symmetric
setting originally proposed in~\cite{JYppt} and have attempted to provide
a systematic analysis of the existence, stability and evolutionary dynamics of
the nonlinear modes that arise in the presence of such a potential
for both self-focusing and self-defocusing nonlinearities. It was
found that all three linear modes generate nonlinear counterparts.
Generally, the defocusing case was found to be more robustly stable
than the focusing one. In the former, two of the branches
were stable for all the values of the frequency considered,
while in the focusing case, all three branches developed instabilities
sufficiently far from the linear limit (although all of them were
spectrally stable close to it). The instabilities could be of
different types, both oscillatory (as for branch 2) and exponential
(as for branch 3) or even of mixed type (as for branch 1). The resulting
 oscillatorily or exponentially unstable dynamics, respectively, led
to the gain overwhelming the dynamics and leading to indefinite growth
in the one or two of the gain peaks of our four-peak potential.

Naturally, there are numerous directions that merit additional investigation.
For instance, and although this would be of less direct relevance
in optics, partial $\mathcal{PT}$ symmetry could be extended to 3 dimensions.
There it would be relevant to appreciate the differences between
potentials that are partially $\mathcal{PT}$ symmetric in one
direction vs. those partially $\mathcal{PT}$ symmetric in two directions.
Another relevant case to explore in the context of the present
mode is that where a $\mathcal{PT}$ phase transition has already
occurred through the collision of the second and third linear
eigenmode considered herein. Exploring the nonlinear modes and
the associated stability in that case would be an interesting
task in its own right. Such studies are presently under consideration
and will be reported in future publications.


\acknowledgments{PGK
gratefully acknowledges the support of NSF-PHY-1602994, the Alexander von
Humboldt Foundation, and the ERC under FP7, Marie Curie Actions, People,
International Research Staff Exchange Scheme (IRSES-605096).  Both authors express their gratitude to Professor Jianke Yang for initiating their interest in this direction and for numerous relevant discussions. }

\bibliographystyle{mdpi}

\begin{thebibliography}{999}

\bibitem{Bender1} Bender, C. M.; Boettcher, S.  Real Spectra in Non-Hermitian Hamiltonians Having $\mathcal{PT}$ Symmetry. {\em Phys. Rev. Lett.} {\bf 1998}, {\em 80}, 5243-5246.

\bibitem{Bender2} Bender, C. M.; Brody, D. C.; Jones, H. F.  Complex Extension of Quantum Mechanics.
    {\em Phys. Rev. Lett.} {\bf 2002}, {\em 89}, 270401-1-270401-4.

\bibitem{Ruter} Ruter, C. E.; Markris, K. G.; El-Ganainy, R.;  Christodoulides, D. N.;  Segev, M.; Kip, D. Observation of parity-time symmetry in optics. {\em Nat. Phys.} {\bf 2010}, {\bf 6}, 192-195.

\bibitem{Peng2014} Peng, B.; Ozdemir, S. K.; Lei, F.;  Monifi, F.;  Gianfreda, M.; Long, G. L.;  Fan, S.; Nori, F.;  Bender, C. M.;  Yang, L.   Parity?time-symmetric whispering-gallery microcavities.   {\em Nat. Phys.} {\bf 2014}, {\em 10}, 394-398.

\bibitem{peng2014b} Peng, B.;  Ozdemir, S. K.; Rotter, S.; Yilmaz, H.;  Liertzer, M.; Monifi, F.; Bender, C. M.; Nori, F.; Yang, L.  
Loss-induced suppression and revival of lasing.  {\em Science} {\bf 2014},  {\em 346}, 328-332.

\bibitem{RevPT} Suchkov, S. V.; Sukhorukov, A. A.; Huang, J.; Dmitriev, S. V.; Lee, C.;  Kivshar, Yu. S.  Nonlinear switching and solitons in PT-symmetric photonic systems. {\em Laser Photonics Rev.} {\bf 2016}, {\em 10}, 177-213.

\bibitem{Konotop}  Konotop, V. V.; Yang, J.; Zezyulin, D. A. Nonlinear waves in $\mathcal{PT}$-symmetric systems.  {\em
  Rev. Mod. Phys.} {\bf 2016}, {\em 88}, 035002-1-035002-65.

\bibitem{Schindler1}  Schindler, J.;  Li, A.;  Zheng, M. C.;  Ellis, F. M.; Kottos,  T.  Experimental study of active LRC circuits with $\mathcal{PT}$ symmetries. {\em Phys. Rev. A} {\bf 2011}, {\em 84}, 040101-1-040101-4.

\bibitem{Schindler2} Schindler, J.;  Lin, Z.;  Lee, J. M.; Ramezani, H.;  Ellis, F. M.; Kottos, T. $\mathcal{PT}$-symmetric electronics.  {\em J. Phys. A: Math. Theor.} {\bf 2012}, {\em 45}, 444029-1-444029-17.

\bibitem{Factor}  Bender, N.; Factor, S.;  Bodyfelt, J. D.; Ramezani, H.;  Christodoulides, D. N.;  Ellis, F. M.; Kottos, T.  Observation of Asymmetric Transport in Structures with Active Nonlinearities. {\em Phys. Rev.  Lett.} {\bf 2013},  {\em 110}, 234101-1-234101-5.

\bibitem{Bender3}  Bender, C. M.;  Berntson, B.; Parker, D.; Samuel, E.  Observation of $\mathcal{PT}$ Phase Transition in a Simple Mechanical System.   {\em Am. J. Phys.} {\bf 2013}, {\em 81}, 173-179.


\bibitem{OptSolPT} Musslimani, Z. H.; Makris, K. G.; El-Ganainy, R.; Christodoulides, D. N.  Optical Solitons in $\mathcal{PT}$ Periodic Potentials. {\em Phys. Rev. Lett.} {\bf 2008}, {\em 100}, 030402-1-030402-4.

\bibitem{WangWang} Wang, H.; Wang, J.  Defect solitons in parity-time periodic potentials. {\em Opt. Express} {\bf 2011}, {\em 19}, 4030-4035.

\bibitem{LuZhang} Lu, Z.; Zhang, Z.  Defect solitons in parity-time symmetric superlattices.  {\em Opt. Express} {\bf 2011}, {\em 19}, 11457-11462.

\bibitem{StabAnPT} Nixon, S.; Ge, L.; Yang, J.  Stability analysis for solitons in $\mathcal{PT}$-symmetric optical lattices.  {\em Phys. Rev. A} {\bf 2012}, {\em 85}, 023822-1-023822-10.

\bibitem{ricardo}  Achilleos, V.;  Kevrekidis, P. G.;  Frantzeskakis, D. J.;  Carretero-Gonz{\'a}lez, R.  Dark solitons and vortices in $\mathcal{PT}$-symmetric nonlinear media: From spontaneous symmetry breaking to nonlinear $\mathcal{PT}$ phase transitions. 
{\em Phys. Rev. A} {\bf 2012}, {\em 86}, 013808-1-013808-7.
 
\bibitem{JYppt} Yang, J.  Partially $\mathcal{PT}$ symmetric optical potentials with all-real spectra and soliton families in multidimensions.  {\em Opt. Lett.} {\bf 2014}, {\em 39}, 1133-1136.

\bibitem{PK4well} Wang, C.; Theocharis,G.; Kevrekidis, P. G.; Whitaker, N.; Law, K. J. H.; Frantzeskakis, D. J.; Malomed, B. A. Two-dimensional paradigm for symmetry breaking:  The nonlinear Schr\"odinger equation with a four-well potential.  {\em Phys. Rev. E} {\bf 2009}, {\em 80}, 046611-1-046611-9.


\end{thebibliography}

\renewcommand\bibname{References}

\end{document}